\documentclass[lettersize,journal]{IEEEtran}
\usepackage{cite}
\usepackage{graphicx}
\usepackage{amsmath,amssymb,amsfonts}
\usepackage[center]{caption}
\usepackage[font=footnotesize]{caption}
\usepackage{xcolor}
\usepackage{algorithm,algorithmicx,algpseudocode}
\usepackage{bm}
\usepackage{mathtools}
\usepackage{nccmath}
\usepackage{enumitem}
\usepackage{array}
\definecolor{Green}{rgb}{0.13, 0.55, 0.13}

\usepackage{amsthm}
\newtheorem{theorem}{Theorem}
\newtheorem{proposition}{Proposition}
\newtheorem{lemma}{Lemma}
\newtheorem{definition}{Definition}

\DeclareMathOperator*{\argmin}{arg\,min}

\title{\huge Semantics-Aware Updates from Remote Energy Harvesting Devices to Interconnected LEO Satellites}

\author{Erfan Delfani, and Nikolaos Pappas, 
		\IEEEmembership{Senior Member, IEEE}.
		\thanks{The authors are with the Department of Computer and Information Science at Linköping University, Sweden, email: \{\texttt{erfan.delfani, nikolaos.pappas\}@liu.se}. This work has been supported in part by the Swedish Research Council (VR), ELLIIT, and the EU (ETHER, 101096526, ELIXIRION, 101120135, and SOVEREIGN, 101131481).}}

\begin{document}

    \setlength{\abovecaptionskip}{3pt}
    \setlength{\belowcaptionskip}{-14pt} 

	\maketitle
	
	\begin{abstract}
       Providing timely and informative data in integrated terrestrial and non-terrestrial networks is critical as data volume grows while the resources available on devices remain limited. To address this, we adopt a semantics-aware approach to optimize the Version Age of Information (VAoI) in a status update system in which a remote Energy Harvesting (EH) Internet of Things (IoT) device samples data and transmits it to a network of interconnected Low Earth Orbit (LEO) satellites for dissemination and utilization. The optimal update policy is derived through stochastic modeling and optimization of the VAoI across the network. The results indicate that this policy reduces the frequency of updates by skipping stale or irrelevant data, significantly improving energy efficiency.
    \end{abstract}


	\section{Introduction}

    The integration of Terrestrial and Non-Terrestrial (T-NT) communication networks has unlocked new opportunities, enabling seamless extended coverage and enhanced remote communication and processing capabilities. Incorporating aerial and space platforms—such as Unmanned Aerial Vehicles (UAVs), High Altitude Platforms (HAPs), and satellites—provides critical infrastructure for connecting remote nodes in rural areas, islands, ships, and airplanes, facilitating global services\cite{ntontinether}. Among NT networks, LEO satellites are experiencing rapid growth, forming a dense web of interconnected nodes around the Earth. These satellites can connect directly to ground devices and gateways, serving as end-users or relays to support global information caching, processing, and management. However, the hardware, software, and energy constraints of satellite nodes make efficient resource management essential for sustaining network performance, particularly when handling large volumes of data. This challenge is further amplified when dealing with time-sensitive data and real-time decision-making from remote IoT devices. In such scenarios, \emph{the communication of timely and informative data plays a crucial role in the performance of the network.}  

    The \emph{semantics-aware communication} is a novel approach that optimizes the generation, transmission, and utilization of fresh and informative data\cite{kountouris2021semantics}. In this framework, semantic attributes such as \emph{timeliness}, \emph{relevance}, and \emph{value} are quantified using metrics like Age of Information (AoI)\cite{kaul2012real}, Age of Incorrect Information (AoII)\cite{maatouk2020age}, and VAoI \cite{yates2021Vage,delfani2024semantics}. AoI measures data freshness but disregards its content. In contrast, AoII and VAoI account for content, with VAoI requiring only minimal knowledge of it. Optimizing these metrics enables integrated T-NT networks to deliver relevant information in a timely manner, minimize outdated or uninformative data, enhance network efficiency, and reduce energy consumption.

    Several studies have examined semantics-aware communication in LEO satellite networks, where the data may be utilized within the LEO network---for example, for distributed processing---or ultimately delivered to a ground destination node via dissemination in the LEO network.
    The works \cite{soret2020latency, chiariotti2022age} investigate AoI and Peak AoI (PAoI) in multihop satellite networks under packet erasure channels and queuing policies. Path selection in dynamic LEO constellations with Inter-Satellite Link (ISL) interruptions is explored in \cite{li2022age}, where AoI is optimized under path and arrival rate constraints. Access control mechanisms for minimizing PAoI in GEO/LEO heterogeneous networks for IoT gateways are examined in \cite{cai2022age}. 
    Optimization of average AoI in integrated networks is addressed in \cite{gao2022non} through a Non-Orthogonal Multiple Access (NOMA)-based two-user scheme, with satellites offering orthogonal access for other users. Protocols for timely dual-hop status updates in Satellite IoT (SIoT) systems, leveraging relaying LEO satellites to enhance PAoI over fading channels, are developed in \cite{huang2023age, jiao2022age}. The work in \cite{xu2022age} introduces an age-optimal delivery protocol for two-hop SIoT links, incorporating data compression, transmission scheduling, and spatial-temporal correlations. In \cite{hong2023age}, an AoII minimization problem is formulated under power, network stability, and freshness constraints in a downlink NOMA-based system. \cite{liao2024information} proposes an LEO IoT architecture with edge intelligence for collaborative task processing, optimizing PAoI while considering terminal energy constraints. Lastly, \cite{badia2025satellite} addresses AoI-minimal scheduling over a finite horizon, with updates from an IoT device delivered to a monitor via orbiting, line-of-sight LEO satellites, accounting for intermittent connectivity.

    The aforementioned studies have focused mainly on information timeliness—AoI and PAoI—while neglecting content-based metrics like VAoI in LEO satellite networks. They also primarily consider individual satellite setups, overlooking the interconnected topology.
    This work addresses these gaps with a semantics-aware approach to handling information from an energy-constrained IoT device to a network of interconnected LEO satellites with two distinct topologies. We analyze the VAoI at LEO nodes and optimize its average across the network by adopting a transmission policy at the IoT device, subject to energy constraints in an EH scenario, where the effective energy use for delivering timely and informative data is targeted.

    \section{System Model}
    \label{Sec_SysModel}

    We consider a system model in which a remote EH IoT device measures and transmits status updates from an information source to a network of $(N+1)$ LEO satellites.
    During a visibility window, the IoT device connects to a satellite, referred to as the Connected Satellite (CS), and transmits updates according to an \emph{update policy} while adhering to the constraints imposed by the harvested energy stored in the device's battery. The update policy, denoted by $\pi$, decides whether the device transmits a fresh update to the CS, thereby consuming energy, or remains idle to conserve energy for future use. This decision, or \emph{action}, is made sequentially in each time slot along a slotted time axis.
    
    \textit{Energy Harvesting:} The device harvests energy from ambient sources and stores it in a battery with capacity $B$. The energy harvesting process follows a Bernoulli distribution with an arrival probability of $\beta$, which is commonly used as a general stochastic model\cite{dong2015near,delfani2024version}. Each transmission to the CS consumes one energy unit and occupies one time slot.
    
    \textit{LEO Network Topologies:} We consider two topologies for the LEO network: a ring topology and a star topology (as depicted in Fig. \ref{fig_SysModel}). 
    In the ring configuration, enabled by permanent ISLs \cite{chaudhry2021laser}, updates received at the CS are disseminated in both directions along a bidirectional ring topology. Nodes are indexed by the set $\mathcal{N_R} = \left\{-\frac{N}{2},-\frac{N}{2}\!+\!1,\cdots,\frac{N}{2}\!-\!1,\frac{N}{2}\right\}$, where $N$ is an even integer. Each node forwards the updates to its neighbors, thereby continuing the propagation throughout the network. We assume that transmissions between neighboring satellites via ISLs are deterministic, occurring error-free and at regular intervals. 
    In the star configuration, the CS multicasts status updates to $N$ neighboring satellite nodes, each one hop away via unreliable ISLs with success probabilities $\rho_n,\ n \in \mathcal{N_S} \setminus \{0\}$, where $\mathcal{N_S} = \left\{0,1,2,\cdots,N\right\}$. This multicast occurs in every time slot. In both topologies, each transmission occurs in one time slot, and each node retains only the most recent data update, discarding any previous ones. 
    
    \textit{VAoI as the Semantic Metric:} The ultimate objective is to develop an update policy that optimizes network performance by delivering timely and informative data while efficiently managing energy, using VAoI as the semantic performance metric. VAoI measures both the timeliness and relevance of information in status update systems, reflecting the number of versions the receiver lags behind the source as new content or versions are generated \cite{yates2021Vage}. By assigning version numbers to new content at the source, the VAoI at a destination node $D$ can be defined as:
    $\Delta(t) \overset{\text{def}}{=} V_S(t) - V_D(t)$,
    where $V_S(t)$ is the version stored at the source, and $V_D(t)$ denotes the version stored at node $D$ at time $t$. We assume that a new version at the source is generated with probability $p_g$ in each time slot, following a Bernoulli distribution.

    \begin{figure*}
		\centering
		\includegraphics[width=0.95\linewidth]{SysModel_New.eps}
		\caption{Status updates from an IoT device to an $(N\!+\!1)$-satellite LEO network: (a) ring, (c) star topology. (b) shows the direct link from the device to the CS.}
        \label{fig_SysModel}
    \end{figure*}

    \section{Average VAoI within the Satellite Network}
    \label{Sec_SysSetup}

    Our objective is to model the average VAoI in the satellite network and optimize it by deriving an optimal update policy. Considering a time horizon $T$, the time-average VAoI of the $n$-th satellite and the average VAoI across the entire network, for a given update policy $\pi$, are defined as follows:
    \begin{align}
        \label{eqn_FiniteHorizonAverageVAoIn}
        \bar{\Delta}_{n,T}^\pi  \overset{\text{def}}{=} \frac{1}{T} \sum_{t=0}^{T-1} \mathbb{E} \left[ \Delta_n^\pi(t) \right] \\
        \label{eqn_FiniteHorizonAverageVAoI}
        \bar{\Delta}_T^\pi \overset{\text{def}}{=} \frac{1}{N+1} \sum_{n \in \mathcal{N}} \bar{\Delta}_{n,T}^\pi
    \end{align}
    where $\Delta_n^\pi(t)$ denotes the VAoI at the $n$-th satellite at time $t$ under the policy $\pi$, and $\mathcal{N}$ is either $\mathcal{N_R}$ or $\mathcal{N_S}$.

    \begin{proposition}
    \label{Prop_VAoIn}
    The VAoI at the $n$-th satellite for the ring and star topologies is given by:
    \begin{align}
        \label{eqn_VAoIn}
        Ring\!: \quad \Delta_{n}(t) &= \zeta_{|n|} + \Delta_0(t-|n|), \quad n \in \mathcal{N_R}, \\
        \label{eqn_VAoIn_Star}
        Star\!: \quad \Delta_{n}(t) &= \zeta_{\mathit{m}_n} + \Delta_0(t-\mathit{m}_n), \quad n \in \mathcal{N_S} \!\setminus\! \{0\},
    \end{align}
    where $\Delta_0(t)$ denotes the VAoI at the Connected Satellite, $\mathit{m}_n$ is a Geometric Random Variable (RV) with parameter $\rho_n$, i.e., $\mathit{m}_n \sim Geom(\rho_n)$, and $\zeta_{k}$ is a Binomial RV with parameters $k$ and $p_g$, i.e., $\zeta_{k} \sim Bin(k, p_g), \ k \in \{0,1,2, \cdots \}$. 
    \end{proposition}

    \begin{proof}
     \textit{Ring:} The VAoI at node $n$ in the ring topology, which is $|n|$ hops away from CS, is given by $\Delta_{n}(t)\!=\!V_S(t) \!-\! V_{n}(t)$, where $V_{n}(t)$ is the stored version at node $n$ at time $t$. The current version at node $n$ is equal to the stored version at the CS in the $|n|$-th slot prior: $V_{n}(t)=V_0(t-|n|).$ Therefore, we can rewrite the VAoI at node $n$ as follows:
    \begin{align}
        \label{eqn_ProofVAoIn}
        \Delta_{n}(t) &= V_S(t) - V_0(t-|n|) \\
        & = \underbrace{V_S(t) - V_S(t-|n|)}_{\zeta_{|n|}} + \underbrace{V_S(t-|n|) - V_0(t-|n|)}_{\Delta_0(t-|n|)} \notag
    \end{align}
    where $\zeta_{|n|} \overset{\text{def}}{=} V_S(t) - V_S(t-|n|)$ counts the number of version generations at the source during the past $|n|$ time slots. The version generation at the source in each time slot follows a Bernoulli distribution with parameter $p_g$. Consequently, the number of version generations during $|n|$ time slots follows a Binomial distribution, i.e., $\zeta_{|n|} \sim Bin(|n|, p_g)$. 
    Eqn. \eqref{eqn_VAoIn} states that the VAoI at node $n$ at time $t$ equals the VAoI at the CS $|n|$ time slots earlier (i.e., at $t - |n|$), plus the number of version changes during the last $|n|$ time slots.

    \textit{Star:} In the star topology, the CS transmits an update to all nodes in each time slot. Upon successful reception, each node retains only the latest version and discards previous ones. Thus, the current VAoI at node $n$ depends on recent transmission outcomes. If the latest transmission succeeds (with probability $\rho_n$), the VAoI at node $n$ equals the VAoI at the CS in the previous slot ($t-1$), plus the number of version generations during the last slot. If it fails (with probability $1-\rho_n$), but the preceding transmission succeeds (with probability $\rho_n$), the VAoI equals the VAoI at the CS two slots ago (i.e., $t-2$), plus the version generations over the past two slots. This occurs with probability $(1-\rho_n)\rho_n$. If both transmissions fail, the node refers to the VAoI at the CS from three slots ago, plus the version generations over the past three slots, and so on. Therefore, the current VAoI at node $n$ equals the VAoI at the CS $\mathit{m}_n$ slots ago, plus the version generations over the last $\mathit{m}_n$ slots, where $\mathit{m}_n$ follows the Geometric distribution $P[\mathit{m}_n = m] = (1-\rho_n)^{m-1}\rho_n, \quad m \in \{1, 2, \cdots\}$.
    \end{proof}

    \begin{lemma}
    \label{Lemma_AvgVAoI}
    Under the update policy $\pi$, the average VAoI at the $n$-th node and across the entire network of LEO satellites over $T$ time slots are, respectively, given by:
    \begin{align}
    \label{eqn_FirstSimpNodeVAoI_Ring}
    &Ring\!: \qquad \bar{\Delta}_{n,T}^\pi \!=\! |n|p_g \!+ \! \mathbb{E} \left[ \frac{1}{T} \! \sum_{t=0}^{T\!-\!1}  \Delta_0^\pi(t\!-\!|n|) \right], \\
        \label{eqn_FirstSimpAvgVAoI}
        &\bar{\Delta}_T^\pi \!=\!  \frac{N(N\!+\!2)}{4(N\!+\!1)}p_g \!+\! \frac{1}{N\!+\!1} \!\!\sum_{n=-\frac{N}{2}}^{\frac{N}{2}} \! \mathbb{E} \left[ \frac{1}{T} \! \sum_{t=0}^{T\!-\!1}  \Delta_0^\pi(t\!-\!|n|) \right]\!. \\
        & Star\!: \notag \\
        \label{eqn_FirstSimpNodeVAoI_Star}
        &\bar{\Delta}_{n,T}^\pi \!=\! \frac{p_g}{\rho_n} \!+ \!  \!\! \sum_{m=1}^{\infty} \!\left\{ \!\rho_n(1\!-\!\rho_n)^{m\!-\!1} \mathbb{E} \! \left[ \frac{1}{T} \! \sum_{t=0}^{T\!-\!1} \Delta_0^\pi(t\!-\!m) \right] \right\}\!, \\
        \label{eqn_FirstSimpAvgVAoI_Star}
        &\bar{\Delta}_T^\pi \!=\! \frac{1}{N+1} \Bigg\{p_g\sum_{n=1}^{N} \frac{1}{\rho_n} \!+ \mathbb{E} \! \left[ \frac{1}{T} \! \sum_{t=0}^{T\!-\!1}  \Delta_0^\pi(t) \right] \\ 
        & \qquad \! + \! \sum_{n=1}^{N} \sum_{m=1}^{\infty} \!\left( \rho_n(1\!-\!\rho_n)^{m-1} \mathbb{E} \! \left[ \frac{1}{T} \! \sum_{t=0}^{T\!-\!1}  \Delta_0^\pi(t\!-\!m) \right] \right) \!\Bigg\}. \notag
    \end{align}
    \end{lemma}

    \begin{proof}
    \textit{Ring:} We can simplify \eqref{eqn_FiniteHorizonAverageVAoIn} using \eqref{eqn_VAoIn}:
    \begin{align}
        \bar{\Delta}_{n,T}^\pi &\!=\! \frac{1}{T} \!\sum_{t=0}^{T\!-\!1} \mathbb{E} \left[ \zeta_{|n|} \!+\! \Delta_0^\pi \left(t\!-\!|n|\right) \right],
        \label{eqn_ProofSimpAvgVAoIn}
    \end{align}
    where by substituting the expected value of $\zeta_{|n|}$, i.e., $\mathbb{E} \left[ \zeta_{|n|} \right] = |n|p_g$, \eqref{eqn_FirstSimpNodeVAoI_Ring} is obtained. By substituting \eqref{eqn_ProofSimpAvgVAoIn} into \eqref{eqn_FiniteHorizonAverageVAoI}, we obtain:
    \begin{align}
        \bar{\Delta}_T^\pi &\!=\!  \frac{1}{N\!+\!1} \!\!\!\sum_{n=-\frac{N}{2}}^{\frac{N}{2}} \!\!\left\{ |n| p_g \!+\! \frac{1}{T} \!\sum_{t=0}^{T\!-\!1} \mathbb{E} \left[ \Delta_0^\pi(t\!-\!|n|) \right] \right\} \\
        & \!=\!  \frac{N(N\!+\!2)}{4(N\!+\!1)}p_g \!+\! \frac{1}{T} \frac{1}{N\!+\!1} \!\sum_{n\!=\!-\!\frac{N}{2}}^{\frac{N}{2}} \! \sum_{t=0}^{T\!-\!1} \mathbb{E} \left[ \Delta_0^\pi(t\!-\!|n|) \right]\!. \notag 
    \end{align} 

    \textit{Star:} By employing the law of total expectation (tower rule), the expected value of VAoI at node $n \in \mathcal{N_S} \!\setminus\! \{0\}$ in \eqref{eqn_VAoIn_Star} is given:
    \begin{align}
        \mathbb{E} \left[ \Delta_n^\pi(t) \right] &\!=\! \mathbb{E} \big[ \mathbb{E} \left[ \Delta_n^\pi(t) \mid \mathit{m}_n \right] \big] \!=\! \mathbb{E} \left[ \mathit{m}_np_g \!+\!  \Delta_0(t\!-\!\mathit{m}_n) \right] \notag \\
        & \!=\! \frac{p_g}{\rho_n} \!+\! \sum_{m=1}^{\infty} P[\mathit{m}_n\!=\!m] \mathbb{E}[\Delta_0(t\!-\!m)], 
    \end{align}
    where by substituting $P[\mathit{m}_n = m] = (1-\rho_n)^{m-1}\rho_n$ and averaging over $t$ and $n$, \eqref{eqn_FirstSimpNodeVAoI_Star} and \eqref{eqn_FirstSimpAvgVAoI_Star} are derived. 
    \end{proof}

    Lemma \ref{Lemma_AvgVAoI} shows that the average VAoI at a single node or across the network depends on the VAoI at the CS and the system parameters, $N$, $p_g$, and $\rho_n$. Thus, optimizing them through the optimal policy $\pi^\ast$ reduces to optimizing the VAoI at the CS. We proceed with the optimization of the Version VAoI across the entire network\footnote{We omit the optimization of the average VAoI at the $n$-th node, as it can be obtained in a similar manner.} in the following section. 

    \section{Optimization Problem}
    \label{sec_ProbFormulation}
    
    We aim to optimize the average VAoI in the network, as represented by \eqref{eqn_FirstSimpAvgVAoI} and \eqref{eqn_FirstSimpAvgVAoI_Star}. We can simplify these two equations to represent them within a unified optimization problem. In the ring topology, node $n$ is $|n|$ hops away from the CS; therefore, its version remains at its initial value (e.g., $0$) during the first $|n|-1$ slots, regardless of the optimal policy. Thus, it is reasonable to include samples starting from $|n|$ in the time average of \eqref{eqn_FirstSimpAvgVAoI}. We also adjust the upper bound of the summation to ensure that each node has the same number of samples, equal to $T^\prime = T-\frac{N}{2}$. This results in:
    \begin{align}
        \label{eqn_InfHorizinApprox}
        \frac{1}{T-\frac{N}{2}} \sum_{t=|n|}^{T-(\frac{N}{2}-|n|)-1}  \Delta_0^\pi(t-|n|) = \frac{1}{T^\prime} \sum_{t=0}^{T^\prime-1}  \Delta_0^\pi(t).
    \end{align}
   
   By substituting \eqref{eqn_InfHorizinApprox} into \eqref{eqn_FirstSimpAvgVAoI}, we can summarize the average VAoI of the ring network as follows:
    \begin{align}
        \label{eqn_FirstSimpAvgVAoIapprox}
        \bar{\Delta}_T^\pi = \frac{N(N+2)}{4(N+1)}p_g + \mathbb{E} \left[ \frac{1}{T^\prime} \sum_{t=0}^{T^\prime-1}  \Delta_0^\pi(t) \right].
    \end{align}

    The average VAoI in the star topology can also be further simplified. For typical ISLs with $\rho_n \geq 0.5$, the term $(1-\rho_n)^{(m-1)}$ in \eqref{eqn_FirstSimpAvgVAoI_Star} decays rapidly to zero as $m$ increases. For values of $m$ greater than a small threshold $M$, typically between 5 and 10, this term can be dropped. Therefore, the upper bound of the summation over $m$ can be set to $M$. Since $T \gg M$, the time average of the VAoI process with small shifts remains constant. That is:
    \begin{align}
        \label{eqn_SummationVAoIapprox_Star}
        \frac{1}{T} \sum_{t=0}^{T-1}  \Delta_0^\pi(t-m) \approx \frac{1}{T} \sum_{t=0}^{T-1}  \Delta_0^\pi(t), \quad m \leq M  \ll T.
    \end{align}

    By substituting \eqref{eqn_SummationVAoIapprox_Star} into \eqref{eqn_FirstSimpAvgVAoI_Star}, we have the average VAoI of the star topology as follows:
    \begin{align}
        \label{eqn_FirstSimpAvgVAoIapprox_Star}
        \bar{\Delta}_T^\pi &\!=\! \frac{p_g\sum_{n=1}^{N} \frac{1}{\rho_n}}{N+1}  + \mathbb{E} \! \left[ \frac{1}{T} \! \sum_{t=0}^{T\!-\!1}  \Delta_0^\pi(t) \right].
    \end{align}

    Now, given the simplified average VAoI for the ring and star topologies in \eqref{eqn_FirstSimpAvgVAoIapprox} and \eqref{eqn_FirstSimpAvgVAoIapprox_Star}, respectively, the optimization of the average VAoI in both networks is formulated as follows:
    \begin{align}
        \bar{\Delta}_T^\ast \overset{\text{def}}{=}& \min_{\pi \in \Pi} \bar{\Delta}_T^\pi, \\
        Ring\!: \ \bar{\Delta}_T^\ast \!=\! \frac{N(N\!+\!2)}{4(N\!+\!1)}p_g &\!+\! \underbrace{\min_{\pi \in \Pi} \mathbb{E} \left[ \frac{1}{T^\prime} \! \sum_{t=0}^{T^\prime-1}  \Delta_0^\pi(t) \right]}_{\bar{\Delta}_0^\ast}, \label{eqn_RingAvgVAo}\\
        Star\!:\ \bar{\Delta}_T^\ast \!=\! \frac{p_g\sum_{n=1}^{N} \frac{1}{\rho_n}}{N+1} &\!+\! \overbrace{\min_{\pi \in \Pi} \mathbb{E} \left[ \frac{1}{T} \! \sum_{t=0}^{T-1}  \Delta_0^\pi(t) \right]},  \label{eqn_StarAvgVAo}
    \end{align}
    where a standard finite-horizon average cost Markov Decision Process (MDP) problem is obtained for both topologies:

    \begin{align}
        \mathcal{P}_1: \quad \bar{\Delta}_0^\ast = \min_{\pi \in \Pi} \frac{1}{\mathcal{T}} \mathbb{E} \left[  \sum_{t=0}^{\mathcal{T}-1}  \Delta_0^\pi(t) \! \mid \! s(0) \right],
    \end{align}
    where $s(0)$ denotes the initial system state, $\Pi$ the set of feasible policies, and $\mathcal{T}$ the MDP time horizon — which is equal to $T'$ for the ring topology and $T$ for the star topology. The problem models a status update system where an EH IoT device monitors a source and transmits updates to the CS node over an error-prone wireless channel, as illustrated in Fig.~\ref{fig_SysModel}b. Each update succeeds with probability $p_s$. The update policy $\pi$ is a sequence of actions $a(t)$, where $a(t) = 1$ denotes a transmission at time $t$, and $a(t) = 0$ indicates idling to conserve energy. The optimal policy $\pi^\ast$ minimizes the average VAoI in problem $\mathcal{P}_1$ and can be computed using dynamic programming~\cite{puterman2014markov}. The solution to this finite-horizon MDP depends on $\mathcal{T}$, the initial state $s(0)$, and time $t$, making it non-stationary. However, for sufficiently large $\mathcal{T}$ (or $T$), $\mathcal{P}_1$ can be approximated by an infinite-horizon average cost MDP, $\mathcal{P}_2$, yielding a stationary policy independent of the initial state—more suitable for analyzing the optimal policy's behavior.

    \begin{align}
        \mathcal{P}_2: \quad \bar{\Delta}_0^\ast = \min_{\pi \in \Pi} \lim_{\mathcal{T} \rightarrow \infty} \frac{1}{\mathcal{T}} \mathbb{E} \left[  \sum_{t=0}^{\mathcal{T}-1}  \Delta_0^\pi(t) \! \mid \! s(0) \right],
    \end{align}
    
    We obtain the solution for MDP problems $\mathcal{P}_1$ and $\mathcal{P}_2$, characterized by state space $S$, action space $A$, transition probability function $P$, and cost function $C$:

    \begin{itemize}[leftmargin=0.15in]
		\item \textit{States}: We define the state vector $s(t)\overset{\text{def}}{=}\left[b(t),\Delta(t)\right]^T \in S$, where $b(t) \in \{0,1,2,\ldots,B\}$ represents the state of the device's battery, and $\Delta(t) \in \{0,1,2,\cdots,\Delta_{\text{max}}\}$ denotes the VAoI at the CS at time $t$. Here, we omit the subscript and superscript of $\Delta_0^\pi(t)$ for simplicity and truncate high VAoI values, as excessively stale data is skipped by the system and thus need not be counted. The resulting state space, $S$, is a finite set.
        
		\item \textit{Actions}: At time $t$, $a(t)=0$ means idle, while $a(t)=1$ means transmitting an update. The action $a(t)$ is restricted to $0$ when $b(t)=0$. 
        
		\item \textit{Transition probabilities}:
		The transition probabilities are presented in Appendix~\ref{sec_TransProb}.
        
		\item \textit{Cost function}: The transition cost function is equal to the VAoI, i.e., $C\big( s(t), a(t),s(t+1)\big)\overset{\text{def}}{=}\Delta(t+1)$.
	\end{itemize}

    \begin{figure*}
		\centering
		\begin{minipage}{.31\textwidth}
			\includegraphics[width=\textwidth]{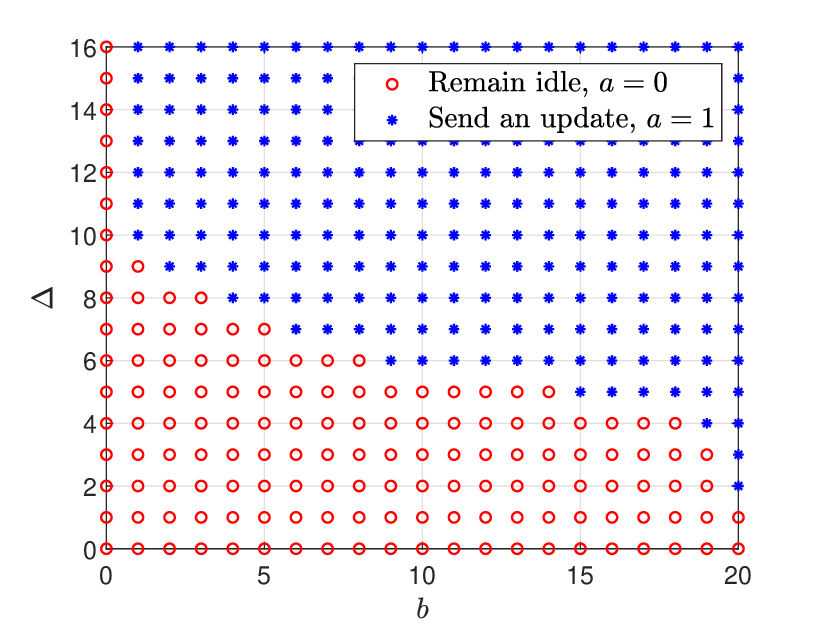}
		  \caption{The structure of the optimal policy for the problem $\mathcal{P}_2$.}
		  \label{fig_OptimalActions}
		\end{minipage}%
        \hspace{8pt}
		\begin{minipage}{.31\textwidth}
			\includegraphics[width=\textwidth]{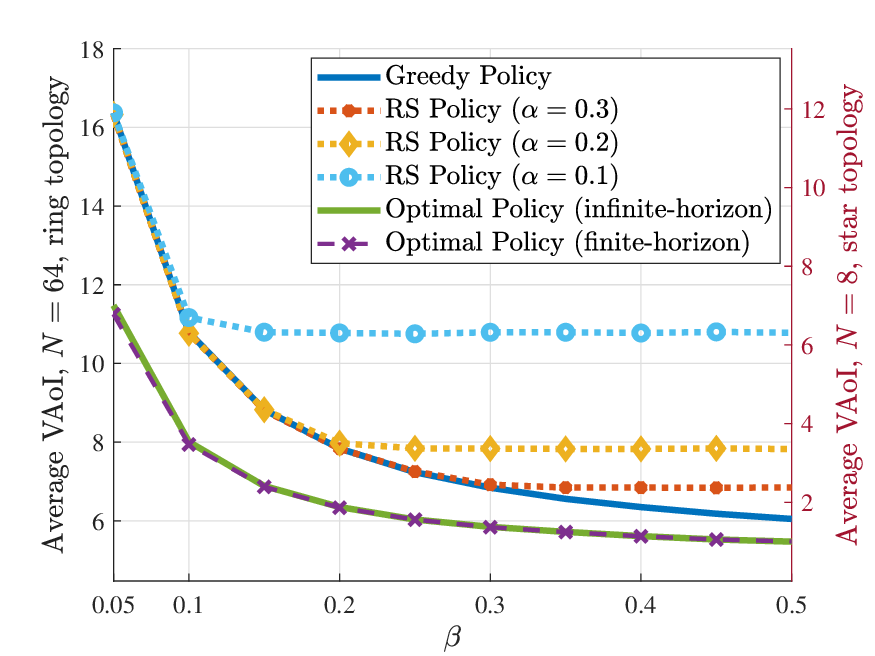}
		  \caption{The average VAoI for various policies vs. $\beta$.}
		  \label{fig_VAoIvsBeta}
		\end{minipage}
        \hspace{8pt}
        \begin{minipage}{.31\textwidth}
			\includegraphics[width=\textwidth]{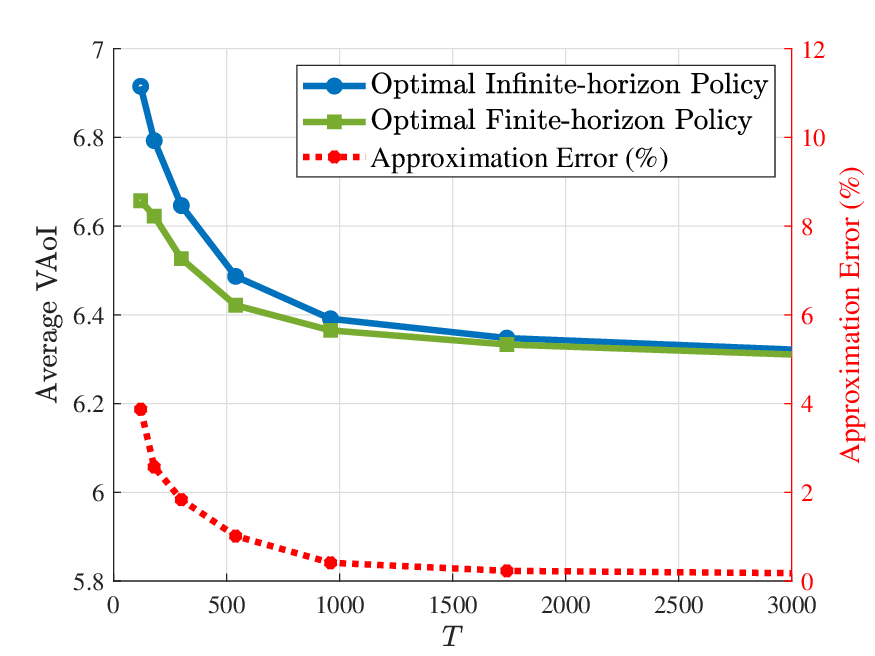}
		  \caption{Impact of $T$ on optimal finite- and infinite-horizon policies ($\beta=0.2$).}
		  \label{fig_ApproxError}
		\end{minipage}
	\end{figure*}

    The existence and structure of the optimal policy for the problem $\mathcal{P}_2$ are analytically investigated in Appendix \ref{sec_AnalyticaOptPolicy}. \vspace{-10pt}

    \section{Numerical Results}
    The optimal policy for $\mathcal{P}_1$ is derived using backward dynamic programming and for $\mathcal{P}_2$ using the Relative Value Iteration Algorithm (RVIA). We compare the performance of the optimal policies with two baselines: the Greedy policy, which transmits an update whenever energy arrives and the battery is not empty, and the Randomized Stationary (RS) policy, which transmits with probability $\alpha$ each time slot, provided the battery is not empty. All simulations use fixed parameters: $p_g = 0.3$, $p_s = 0.5$, $B = 20$, $\Delta_{\text{max}} = 30$, and $\rho_n = 0.7$ for all $n$, unless stated otherwise. Expected values are averaged over $4000$ Monte Carlo iterations.

    \subsection{The Structure of the Optimal Policy}

    Fig. \ref{fig_OptimalActions} shows the optimal policy for $\mathcal{P}_2$ with an energy arrival probability $\beta=0.1$. Red circles denote idle actions ($a=0$), and blue asterisks indicate updates ($a=1$). This policy follows a threshold-based structure: for each battery state $b$, the device remains idle until the VAoI at the CS exceeds a certain threshold, which triggers updates. This reflects a key aspect of semantics-aware communication: \emph{conserving energy for later usage when it is most needed}. In energy-scarce settings, this prevents early battery depletion at low VAoI, avoiding long update gaps. Delaying depletion until VAoI is reasonably high mitigates excessive growth.

    \subsection{The Impact of Energy Arrival Probability $(\beta)$}

    Fig. \ref{fig_VAoIvsBeta} shows the average VAoI for both topologies versus energy arrival probability $\beta$, under optimal, Greedy, and RS policies, for $\alpha = 0.1$, $0.2$, and $0.3$. A large horizon $T = 3000$ is used, where finite- and infinite-horizon policies yield similar results, with the latter slightly better. As seen in \eqref{eqn_RingAvgVAo} and \eqref{eqn_StarAvgVAo}, the average VAoI differs between the ring and star topologies by a constant offset, reflected in the separate $y$-axes. As $\beta$ increases, average VAoI decreases, with the optimal policies performing best using optimal update thresholds. The performance gap between optimal and Greedy narrows with higher energy availability, approaching the \emph{always update} policy. However, this gap widens under low $\beta$, making optimal actions crucial for a fresher, more informative system.

    In this simulation, the RS policy performs worse than the Greedy policy. Notably, the Greedy policy is a special case of the RS policy with $\alpha = 1$, so increasing $\alpha$ causes RS to converge toward Greedy, as shown in Fig.~\ref{fig_VAoIvsBeta}. When the energy arrival probability is high, energy is more often available. The Greedy policy uses all available energy, while the RS policy with $\alpha < 1$ sometimes skips updates, allowing energy to accumulate. As a result, Greedy yields more updates and RS fewer, leading to a worse average VAoI for RS. In contrast, when the energy arrival probability is low, both policies deplete the battery in a similar manner---Greedy by transmitting at the first opportunity and RS at a random slot---resulting in the same number of updates and similar performance.

    \textit{Remark:} 
    An important result is that to maintain a target average VAoI—--such as $8$ for the ring topology or $3.5$ for the star—--the optimal policy requires an energy arrival rate of $0.1$, half that of the Greedy policy ($0.2$). This shows that \emph{a semantics-aware update policy can significantly reduce the energy consumption by $50\%$}. The reduction stems from fewer updates, leading to less satellite dissemination, significantly improving energy efficiency and extending system lifetime.

    \subsection{The Impact of the Time Horizon $(T)$}
    Section \ref{sec_ProbFormulation} presented the optimization problem for both finite- and infinite-horizon cases, where the infinite-horizon problem $\mathcal{P}_2$ approximates the finite-horizon problem $\mathcal{P}_1$ as $T$ increases. Here, we analyze how $T$ affects the approximation. Fig. \ref{fig_ApproxError} shows the average VAoI for a ring network with $N=64$ under the optimal policies of both problems, with the approximation error (in red) shown as a percentage on the right $y$-axis. The curves confirm that as $T$ increases, the error becomes negligible: it drops below $4\%$ for $T \geq 2N$ and below $1\%$ for $T \geq 10N$.

    \section{Conclusion}
    We analyzed and optimized the VAoI to ensure the timely transmission of informative updates from a ground-based IoT device to a network of interconnected LEO satellites. Within this network, updates are disseminated among nodes arranged in a ring or star topology. By formulating and optimizing the VAoI across the LEO network, the proposed policy reduces the transmission of stale and irrelevant updates, thereby enhancing energy efficiency.

    \bibliographystyle{IEEEtran}
    \bibliography{Refs}

    \newpage

    \appendices

    \section{Transition Probabilities}
    \label{sec_TransProb}
    The transition probabilities between the system states are presented by introducing the following Bernoulli processes: the energy arrival process, $\mathit{\textit{e}}(t)$, the channel success process, $\mathit{\textit{c}}(t)$, and the version generation process, $\mathit{\textit{z}}(t)$, given by:
	\begin{align}
		\label{BernoulliProcesses}
		\begin{matrix}
			\mathit{\textit{e}}(t) \!=\!
			\begin{cases}
				1 & \text{w.p. } \beta, \\
				0 & \text{w.p. } \bar{\beta}, \\
			\end{cases} & 
			\mathit{\textit{c}}(t) \!=\! 
			\begin{cases}
				1 & \text{w.p. } p_s, \\
				0 & \text{w.p. } \bar{p}_s, \\
			\end{cases} \\
			\mathit{\textit{z}}(t) \!=\! 
			\begin{cases}
				1 & \text{w.p. } p_g, \\
				0 & \text{w.p. } \bar{p}_g. \\
			\end{cases}
		\end{matrix}
	\end{align}
    where $\bar{\beta}\overset{\text{def}}{=}1-\beta$, $\bar{p}_s\overset{\text{def}}{=}1-p_s$, and $\bar{p}_g\overset{\text{def}}{=}1-p_g$. 
	We can now describe the evolution of the states based on the explanation provided in Section \ref{Sec_SysModel}:
	\begin{align}
		b(t\!+\!1) &\!=\! \min \left\{ b(t) \!+\! \mathit{\textit{e}}(t) \!-\! a(t) ,B \right\}. \\
		\Delta(t\!+\!1) 
		& \!=\! \begin{cases}
			\mathit{\textit{z}}(t), \quad  a(t)\!=\!1 \text{ and } \mathit{\textit{c}}(t)\!=\!1, \\
			\min \! \left\{ \Delta(t) \!+\! \mathit{\textit{z}}(t) ,\Delta_{\text{max}} \right\}\!, \ \text{otherwise.}
		\end{cases}
	\end{align}
	
	Given the following equation: 
		\begin{align}
			\label{TransProb_Eqn}
			P\left[s(t\!+\!1)| s(t),a(t)\right]\!&=\!P\left[b(t\!+\!1)|b(t),a(t)\right]  \\
			&\times  P\left[\Delta(t\!+\!1)|b(t),\Delta(t),a(t)\right], \notag
		\end{align}
    the transition probabilities are determined by the following equations:
	\begin{align}
		P&\left[b(t\!+\!1)\big|b(t),a(t)\right] \\
        &\!=\! 
		\begin{cases}
			\beta & a(t)\!=\!0,\  b(t\!+\!1)\!=\!b(t)\!+\!1, \ b(t)\!<\!B,\\
			\bar{\beta} & a(t)\!=\!0,\  b(t\!+\!1)\!=\!b(t), \ b(t)\!<\!B,\\
            1 & a(t)\!=\!0, \ b(t\!+\!1)\!=\!b(t)\!=\!B, \\
			\beta & a(t)\!=\!1,\  b(t\!+\!1)\!=\!b(t),\\
			\bar{\beta} & a(t)\!=\!1,\  b(t\!+\!1)\!=\!b(t)\!-\!1.
		\end{cases} \notag
	\end{align}
	\begin{align}
		P&\left[\Delta(t\!+\!1) \big|b(t),\Delta(t),a(t)\right] \\
		&\!=\! 
		\begin{cases}
			p_g & a(t)\!=\!0,\  \Delta(t\!+\!1)\!=\!\Delta(t)\!+\!1,\ \Delta(t)\!\!<\!\!\Delta_{\text{max}},\\
			\bar{p}_g & a(t)\!=\!0,\  \Delta(t\!+\!1)\!=\!\Delta(t),\ \Delta(t)\!\!<\!\!\Delta_{\text{max}},\\
            1 & a(t)\!=\!0,\  \Delta(t\!+\!1)\!=\!\Delta(t)\!=\!\Delta_{\text{max}},\\
			p_g\bar{p}_s & a(t)\!=\!1,\  \Delta(t\!+\!1)\!=\!\Delta(t)\!+\!1,\ \Delta(t)\!\!<\!\!\Delta_{\text{max}},\\
			\bar{p}_g\bar{p}_s & a(t)\!=\!1,\  \Delta(t\!+\!1)\!=\!\Delta(t),\ \Delta(t)\!\!<\!\!\Delta_{\text{max}},\\
            \bar{p}_s & a(t)\!=\!1,\ \Delta(t\!+\!1)\!=\!\Delta(t)\!=\!\Delta_{\text{max}},\\
			p_gp_s & a(t)\!=\!1,\  \Delta(t\!+\!1)\!=\!1,\\
			\bar{p}_gp_s & a(t)\!=\!1,\  \Delta(t\!+\!1)\!=\!0.
		\end{cases} \notag
	\end{align}

    \section{The Optimal Policy Structure for Problem $\mathcal{P}_2$}
    \label{sec_AnalyticaOptPolicy}

    In this section, we first prove the existence of the optimal policy for the MDP problem $\mathcal{P}_2$ and its adherence to Bellman's equation. We then demonstrate the threshold-based structure using VIA and mathematical induction.

    \begin{definition}
		An MDP is considered weakly accessible if its state space can be partitioned into two subsets, $S_t$ and $S_c$, where all states in $S_t$ are transient under every stationary policy, and every state $s^\prime$ in $S_c$ is reachable from any other state $s$ in $S_c$ under some stationary policy.
	\end{definition}
	
	\begin{proposition}
		\label{WeaklyAccessProp}
		The MDP problem $\mathcal{P}_2$ is weakly accessible.
	\end{proposition}
	
	\begin{proof}
		We show that any state $s^\prime = \left(b^\prime, \Delta^\prime\right) \in S$ can be reached from any other state $s = \left(b, \Delta\right) \in S$ under a stationary stochastic policy $\pi$, in which the action $a \in \{0,1\}$ is chosen at each state with a positive probability. If $b^\prime < b$, the state $b^\prime$ is reachable from $b$ with positive probability (w.p.p.) by repeatedly selecting action $a = 1$ for $(b - b^\prime)$ time slots. If $b^\prime \geq b$, the state $b^\prime$ is reachable w.p.p. by selecting action $a = 0$ for $(b^\prime - b)$ slots. Once the system reaches state $b^\prime$, the battery level can remain unchanged w.p.p., regardless of the subsequent actions. Therefore, we treat the battery state as $b^\prime$ for the remainder of the proof. Similarly, the state $\Delta^\prime < \Delta$ can be reached from $\Delta$ w.p.p. by first executing action $a = 1$ for one time slot, followed by action $a = 0$ for $\Delta^\prime$ slots. Conversely, if $\Delta^\prime \geq \Delta$, the state $\Delta^\prime$ is reachable by selecting action $a = 0$ for $(\Delta^\prime - \Delta)$ slots. 
	\end{proof}
	
	\begin{proposition}
		In the MDP problem $\mathcal{P}_2$, the optimal average cost $J^\ast$ achieved by an optimal policy $\pi^\ast$ is the same for all initial states. This optimal cost satisfies Bellman’s equation:
		\vspace{-5pt}
		\begin{equation}
			\label{Bellman_eqn}
			J^\ast\!+\!V(s)\!=\!\min_{a\in\left\{0,1\right\}}{\!\bigg\{C(s,a)+\!\sum_{s^\prime\in \mathcal{S}}{P\left(s^\prime \big | s,a\right)\!V(s^\prime)}\!\bigg\}},
			\vspace{-4pt}
		\end{equation}
		\begin{equation}
			\label{OptimalAction_eqn}
			\pi^\ast(s) \in  \argmin_{a\in\left\{0,1\right\}}{\!\bigg\{C(s,a)+\!\sum_{s^\prime\in \mathcal{S}}{P\left(s^\prime \big | s,a\right)\!V(s^\prime)}\!\bigg\}},
			\vspace{-4pt}
		\end{equation}
		where $V(s)$ denotes the value function of the MDP, $P(s' \mid s,a)$ is the transition function, and $C(s,a)$ represents the average cost per slot, which is determined by the transition costs:
		\begin{equation}
			\label{AvgCost_eqn}
			C(s,a) = \sum_{s^\prime\in \mathcal{S}} {P\left(s^\prime \big | s,a\right) C\left(s,a,s^\prime\right)}, 
		\end{equation}
		with $C(s,a,s^\prime)=\Delta^\prime$.
	\end{proposition}
	
	\begin{proof}
		As stated in Proposition \ref{WeaklyAccessProp}, the MDP is weakly accessible. Consequently, according to Proposition 4.2.3 in \cite{bertsekas2011dynamic}, the optimal average cost remains the same for all initial states. Furthermore, Proposition 4.2.6 in \cite{bertsekas2011dynamic} guarantees the existence of an optimal policy. Additionally, Proposition 4.2.1 in \cite{bertsekas2011dynamic} demonstrates that by determining $J^*$ and $V(s)$ that satisfy \eqref{Bellman_eqn}, the optimal policy can be derived using \eqref{OptimalAction_eqn}.
	\end{proof}
		
	\begin{definition}
		Assume that for each $b$, there exists a value $\Delta_{T}(b) > 0$ such that the action $\pi(b, \Delta)$ is equal to 1 when $\Delta \geq \Delta_{T}(b)$, and 0 otherwise. Under these conditions, $\pi$ is referred to as a threshold policy.
	\end{definition}
	
	\begin{theorem}
		The optimal policy for the MDP problem $\mathcal{P}_2$ is a threshold policy.
	\end{theorem}

    \begin{proof}
		The Bellman equation for the state $s = (b, \Delta)$ is expressed as follows:
		
		\begin{align}
			J^\ast\!+\!V(s)\!&=\!\min_{a\in\left\{0,1\right\}} {\bigg\{\underbrace{\sum_{s^\prime\in S} P\Big[s^\prime \big|s,a\Big] \Big( \Delta^\prime \!+\! V(s^\prime) \Big)}_{\overset{\text{def}}{=}\ Q(a)}\bigg\}} \\
			a^\ast(s)&=\argmin_{a\in\left\{0,1\right\}}{Q(a)}=
			\begin{cases}
				0, & DV(s)\! \geq \!0,\\
				1, & DV(s)\! < \!0,\\
			\end{cases}
		\end{align}
		where $DV(s)\!\overset{\text{def}}{=}\!V^1(s)\!-\!V^0(s)$, $V^0(s)\overset{\text{def}}{=}Q\left(a\!=\!0\right)$, and $V^1(s)\overset{\text{def}}{=}Q\left(a\!=\!1\right)$.
		
		As observed, the optimal action $a^\ast(s)$ depends on the sign of $DV(s)$. When $b=0$, the action $a=0$ is forced, resulting in $DV(s)=0$. For cases where $b>0$, we have:
		\begin{align}
			\label{eqn_DV_1rate}
				V^0(s) &\!=\! \sum_{s^\prime\in S} P\big[s^\prime \big|s,a\!=\!0\big] \Big( \Delta^\prime \!+\! V(s^\prime) \Big) \\ 
                &\!=\! \sum_{\substack{z \in \{0,1\} \\ e \in \{0,1\}}} \Big\{ \big(\Delta\!+\!z\big) \!+\! V(b\!+\!e,\Delta\!+\!z) \Big\} P_e P_z, \notag \\
				V^1(s) &\!=\! \sum_{s^\prime\in S} P\big[s^\prime \big|s,a\!=\!1\big] \Big( \Delta^\prime \!+\! V(s^\prime) \Big) \\
				& \!=\! \sum_{\substack{z \in \{0,1\} \\ e \in \{0,1\}}} \Big\{ \bar{p}_s\big[\left(\Delta\!+\!z\right) \!+\! V(b\!+\!e\!-\!1,\Delta\!+\!z)\big]  \notag \\ 
                & \qquad \qquad + p_s \big[ z \!+\! V(b\!+\!e\!-\!1,z) \big] \Big\} P_e P_z. \notag 
		\end{align}
		where $P_e\!\overset{\text{def}}{=}\!P\left[\mathit{\textit{e}}(t)\!=\!e\right]$, and $P_z\!\overset{\text{def}}{=}\!P\left[\mathit{\textit{z}}(t)\!=\!z\right]$, as specified in \eqref{BernoulliProcesses}.
		In the following, we show that $DV(s) = DV(b, \Delta)$ is a non-increasing (or decreasing) function of $\Delta$. Specifically, for $\Delta^- \leq \Delta^+$, we demonstrate that $DV(b, \Delta^+) \leq DV(b, \Delta^-)$, or equivalently, $DV(b, \Delta^+) - DV(b, \Delta^-) \leq 0$. This behavior leads to the threshold policy, as a negative value of $DV(s)$ for a particular $\Delta_T$ implies that $DV(s)$ will also be negative for any $\Delta \geq \Delta_T$, and thus the optimal action remains $1$. By simplifying $DV(b, \Delta^+)$ and $DV(b, \Delta^-)$ using equation \eqref{eqn_DV_1rate}, we derive the following equations:
		\begin{align*}
				DV&(b,\Delta\!^+) \!-\! DV(b,\Delta\!^-) \\ 
                & \!=\! V^1(b,\Delta\!^+) \!-\! V^1(b,\Delta\!^-) \!-\! \big[V^0(b,\Delta\!^+) \!-\! V^0(b,\Delta\!^-)\big]  \\
				&= \sum_{z, e} \bigg\{ \overbrace{p_s\left(\Delta\!^- - \Delta\!^+\right)}^{\leq 0} \\
                & \quad +\bar{p}_s\Big[V(b\!+\!e\!-\!1,\Delta\!^+\!+\!z) - V(b\!+\!e\!-\!1,\Delta\!^-\!+\!z)\Big]  \\
				& \quad - \Big[ V(b\!+\!e,\Delta\!^+\!+\!z) - V(b\!+\!e,\Delta\!^-\!+\!z)\Big] \bigg\} P_e P_z.
		\end{align*}
		
		Therefore, to verify the inequality $DV(b, \Delta\!^+) - DV(b, \Delta\!^-) \leq 0$, it is enough to demonstrate that 
		\begin{align}
			\bar{p}_s \big[V(b-1,\Delta\!^+) - V(b-1,\Delta\!^-) \big] \\
            - \big[ V(b,\Delta\!^+) - V(b,\Delta\!^-) \big] \!\leq\! 0, \notag
		\end{align}
		for $b>0$ and $\Delta\!^- \leq \Delta\!^+$. 
		To continue with the proof, we apply the VIA along with mathematical induction. The VIA converges to the value function of Bellman's equation, independent of the initial value of $V_0(s)$. Specifically, we have $\lim_{k\rightarrow\infty}{V_k(s)}=V(s),\ \forall s\in S$.
		\begin{align}
			\label{VIA_Iter_k}
			V_{k+1}(s)\!=\!\min_{a\in\left\{0,1\right\}} {\bigg\{\sum_{s^\prime\in S} P\Big[s^\prime \big|s,a\Big] \Big( \Delta^\prime \!+\! V_k(s^\prime) \Big)\bigg\}}.
		\end{align}
		
		Thus, it is enough to demonstrate the following inequality for all values of $k \in \{0,1,2,\cdots\}$:
		\begin{align}
			\label{eqn_DV_diff_iter_k}
			\bar{p}_s \big[V_k(b-1,\Delta\!^+) - V_k(b-1,\Delta\!^-) \big] \\ 
                - \big[ V_k(b,\Delta\!^+) - V_k(b,\Delta\!^-) \big] \leq 0. \notag
		\end{align}
		
		Assuming that $V_0(s) = 0$ for all $s \in S$, equation \eqref{eqn_DV_diff_iter_k} holds for $k = 0$. Next, by extending the assumption in \eqref{eqn_DV_diff_iter_k} for $k > 0$, we seek to demonstrate its validity for $k+1$, i.e.,
		\begin{align}
			\label{eqn_DV_diff_iter_kp1}
			\bar{p}_s \big[V_{k+1}(b\!-\!1,\Delta\!^+) - V_{k+1}(b\!-\!1,\Delta\!^-) \big] \\
            - \big[ V_{k+1}(b,\Delta\!^+) - V_{k+1}(b,\Delta\!^-) \big] \leq 0. \notag
		\end{align}
		
		The VIA equation in \eqref{VIA_Iter_k} is expressed as $V_{k+1}(s) = \min \{V^0_{k+1}(s), V^1_{k+1}(s)\}$, where the following definitions apply:
		\begin{align}
				V^0_{k+1}(s)\overset{\text{def}}{=}\sum_{s^\prime\in S} P\big[s^\prime \big|s,a=0\big] \big( \Delta^\prime \!+\! V_k(s^\prime) \big), \\
				V^1_{k+1}(s)\overset{\text{def}}{=}\sum_{s^\prime\in S} P\big[s^\prime \big|s,a=1\big] \big( \Delta^\prime \!+\! V_k(s^\prime) \big),
		\end{align}
		where 
		\begin{align}
			\label{eqn_RVI_VoV1}
				V^0_{k+1}(s) \!=\! \sum_{z, e} \Big\{ \big(\Delta\!+\!z\big) \!+\! V_k(b\!+\!e,\Delta\!+\!z) \Big\} P_e P_z, \\
				V^1_{k+1}(s)\!=\! \sum_{z, e} \Big\{ \bar{p}_s\big[\left(\Delta\!+\!z\right) \!+\! V_k(b\!+\!e\!-\!1,\Delta\!+\!z)\big] \notag \\ 
                + p_s \big[ z \!+\! V_k(b\!+\!e\!-\!1,z) \big] \Big\} P_e P_z.
		\end{align}
		
		The inequality in \eqref{eqn_DV_diff_iter_kp1} can be simplified further:
		\begin{align}
			\label{eqn_DV_diff_iter_kp1_LastIneq}
			&\bar{p}_s \Big[ \min \{V^0_{k+1}(b\!-\!1,\Delta\!^+),V^1_{k+1}(b\!-\!1,\Delta\!^+)\}  \notag \\ 
            &-\! \min \{V^0_{k+1}(b\!-\!1,\Delta\!^-),V^1_{k+1}(b\!-\!1,\Delta\!^-)\} \Big] \notag \\ 
			&- \Big[ \min \{V^0_{k+1}(b,\Delta\!^+),V^1_{k+1}(b,\Delta\!^+)\} \notag \\ 
            &- \min \{V^0_{k+1}(b,\Delta\!^-),V^1_{k+1}(b,\Delta\!^-)\} \Big] \leq 0 .
		\end{align}
		
		We examine four distinct cases in order to proceed with the proof of \eqref{eqn_DV_diff_iter_kp1_LastIneq}.
			\begin{align*}
				\begin{array}{l}
					\text{\normalsize Case 1. } 
					\begin{cases}
						V^0_{k+1}(b\!-\!1,\Delta\!^-) \leq V^1_{k+1}(b\!-\!1,\Delta\!^-), \\ 
						V^0_{k+1}(b,\Delta\!^+) \leq V^1_{k+1}(b,\Delta\!^+).
					\end{cases} \\
					\text{\normalsize Case 2. } 
					\begin{cases}
						V^0_{k+1}(b\!-\!1,\Delta\!^-) \leq V^1_{k+1}(b\!-\!1,\Delta\!^-), \\ 
						V^0_{k+1}(b,\Delta\!^+) > V^1_{k+1}(b,\Delta\!^+).
					\end{cases} \\
					\text{\normalsize Case 3. } 
					\begin{cases} 
						V^0_{k+1}(b\!-\!1,\Delta\!^-) > V^1_{k+1}(b\!-\!1,\Delta\!^-), \\ 
						V^0_{k+1}(b,\Delta\!^+) \leq V^1_{k+1}(b,\Delta\!^+).
					\end{cases} \\
					\text{\normalsize Case 4. } 
					\begin{cases}
						V^0_{k+1}(b\!-\!1,\Delta\!^-) > V^1_{k+1}(b\!-\!1,\Delta\!^-), \\
						V^0_{k+1}(b,\Delta\!^+) > V^1_{k+1}(b,\Delta\!^+).
					\end{cases}
				\end{array}
			\end{align*}
		
		We prove the inequality \eqref{eqn_DV_diff_iter_kp1_LastIneq} for case 1. A similar method can be applied to prove the remaining cases.
		
		\textit{\textbf{Case 1.}} $V^0_{k+1}(b\!-\!1,\Delta\!^-) \leq V^1_{k+1}(b\!-\!1,\Delta\!^-)$ and $V^0_{k+1}(b,\Delta\!^+) \leq V^1_{k+1}(b,\Delta\!^+)$. In this case, equation \eqref{eqn_DV_diff_iter_kp1_LastIneq} is further simplified:
		\begin{align}
			&\bar{p}_s \Big[ V^0_{k+1}(b\!-\!1,\Delta\!^+) \!-\! V^0_{k+1}(b\!-\!1,\Delta\!^-) \Big] \notag \\ & + \underbrace{ \bar{p}_s \min \{0,V^1_{k+1}(b\!-\!1,\Delta\!^+)\!-\!V^0_{k+1}(b\!-\!1,\Delta\!^+)\} }_{\leq 0}  \notag\\
			& - \Big[ V^0_{k+1}(b,\Delta\!^+) \!-\! V^0_{k+1}(b,\Delta\!^-)\Big] \notag \\
            & + \underbrace{ \min \{0,V^1_{k+1}(b,\Delta\!^-)\!-\!V^0_{k+1}(b,\Delta\!^-)\} }_{\leq 0}  \leq 0,
		\end{align}
		where we have applied the identity $\min\left\{x,y\right\} = x + \min\left\{0, y - x\right\}$. Since the second and last terms are non-positive, it is sufficient to demonstrate that:
		\begin{align}
			\bar{p}_s \Big[ V^0_{k+1}(b\!-\!1,\Delta\!^+) - V^0_{k+1}(b\!-\!1,\Delta\!^-) \Big] -\\ 
            \Big[ V^0_{k+1}(b,\Delta\!^+) - V^0_{k+1}(b,\Delta\!^-)\Big]    \leq 0.
		\end{align}
		
		According to \eqref{eqn_RVI_VoV1}, it can be written as follows:
		\begin{align}
				&\bar{p}_s \sum_{z, e} \Big\{ \big(\Delta\!^+\!-\!\Delta\!^-\big) \!+\! V_k(b\!+\!e\!-1,\Delta\!^+\!+\!z) \notag \\
                &\qquad \qquad - V_k(b\!+\!e\!-1,\Delta\!^-\!+\!z) \Big\} P_e P_z \notag \\
				&- \sum_{z, e} \Big\{ \big(\Delta\!^+\!-\!\Delta\!^-\big) \!+\! V_k(b\!+\!e\!,\Delta\!^+\!+\!z) \notag \\ 
                &\qquad \qquad - V_k(b\!+\!e\!,\Delta\!^-\!+\!z) \Big\} P_e P_z \leq 0 \\
				&\Leftrightarrow \sum_{z, e} \Big\{ \overbrace{(1-\bar{p}_s)(\Delta\!^--\Delta\!^+)}^{\leq 0} \notag \\
                &\quad + \bar{p}_s \big[ V_k(b\!+\!e\!-1,\Delta\!^+\!+\!z) - V_k(b\!+\!e\!-1,\Delta\!^-\!+\!z) \big] \notag \\
				&\quad - \big[ V_k(b\!+\!e\!,\Delta\!^+\!+\!z) \!-\! V_k(b\!+\!e\!,\Delta\!^-\!+\!z) \big] \Big\} P_e P_z \leq 0 
		\end{align}
		where the first term in the summation is negative because $\Delta\!^- \leq \Delta\!^+$ and $1-\bar{p}_s = p_s > 0$. The other terms are also negative, based on the induction hypothesis given in \eqref{eqn_DV_diff_iter_k}, completing the proof.
	\end{proof}
    
\end{document}